\documentclass[12pt]{iopart}

\usepackage{cite}

\usepackage{subeqnarray}
\usepackage{cases}
 
\usepackage{hyperref}
\hypersetup{
		colorlinks=true,
		linkcolor=blue,
		citecolor=blue,
           urlcolor=blue
}

\usepackage{graphicx}

\usepackage{mathrsfs}

\usepackage{siunitx}
\begin{document}
\nocite{*}

\title{Spontaneous decay induced quantum dynamics in  Rydberg blockaded  $\Lambda$-type atoms}

\author{Chang Qiao$^{1,2,3,4*}$ and Wenxian Zhang$^{5}$}
\address{1 International Center for Quantum Materials, School of Physics, Peking University, Beijing 100871, China}
\address{2 Collaborative Innovation Center of Quantum Matter, Beijing 100871, China}
\address{3 Hefei National Laboratory for Physical Sciences at the Microscale and Department of Modern Physics, University of Science and Technology of China, Hefei, Anhui 230026, China}
\address{4 CAS Center for Excellence and Synergetic Innovation Center in Quantum Information and Quantum Physics, University of Science and Technology of China, Shanghai 201315, China}
\address{5 Key Laboratory of Artificial Micro- and Nano-structures of Ministry of Education and School of Physics and Technology,
Wuhan University, Wuhan 430072, China}
\address{$*$ Author to whom any correspondence should be addressed.}

\ead{qch19@mail.ustc.edu.cn}

\vspace{10pt}
\begin{indented}
\item[]October 2021
\end{indented}

\begin{abstract}
Strongly Rydberg-blockaded two-level atoms form a Rydberg superatom, which is excited only to a collective symmetrical Dicke state. However, emerging often in the alkali-earth atoms, the spontaneous decay from the Rydberg state to an additional pooling state  renders the ensemble no longer a closed superatom.  Herein we present a computationally-efficient model to characterize the interaction between a fully Rydberg-blockaded  ensemble of $N$ $\Lambda$-type three-level atoms  and a strong probe light field in a coherent state.  The model enables us to achieve  a decomposition of the coupled dynamics in the strong field limit, which significantly reduces the complexity of computing the $N$-body system evolution and paves a way for practical analysis in experiments. A quasi-steady-state power spectrum with multiple sidebands is found in the scattered field.  The relative heights of the  sidebands show a time dependence determined by the atomic relaxation, which illuminates potential applications of using the system in information transfer of quantum networks. With negligible dissipative flipping  to the unsymmetrical  states, the atomic relaxation time indicating a linearly increasing pooling state fraction is derived analytically as a function of the atom number.   
\end{abstract}

\section{Introduction}
Quantum dynamics between photons and  atoms contributes remarkably to the  developments in quantum optics, quantum simulation and quantum information processing. For instance, the light fields scattered by dense atomic ensembles show intriguing nonlinear properties \cite{Dicke, Scully}, which also emerge in the strongly-correlated Rydberg atoms.  Due to the dipole blockade effect \cite{SuperatomTheory},  the multiple Rydberg excitations are dramatically suppressed.  The fully Rydberg-blockaded atoms are excited only to a collective Dicke state,  performing as a  superatom with a strongly enhanced  transition dipole moment \cite{SuperatomEvidence}.    This feature offers great opportunities for studying the free-space light-matter interactions \cite{propagatingQED2,QEDSuperatom1, QEDSuperatom2,singlephotonemitterLukin} as well as demonstrating potential applications in  quantum networks \cite{QuantumNetworks,LukinRMP}. 

Different dissipation processes have been extensively investigated  in the studies of the Rydberg blockade physics. For example, the photonic dissipations influence the photon-photon interactions mediated by the Rydberg blockaded atoms under the condition of electromagnetically induced transparency  \cite{photonicDissipative1,photonicDissipative2}. On the other hand, while the long-term performance of the Rydberg dressed ensemble  attracts lot of research interest \cite{RDressingTPfau,RydbergDressing1,RydbergDressing2,RydbergDressing3,RydbergDressingMOTMPA},  the atomic dissipation caused by the spontaneous decay from the Rydberg state to an additional pooling state has long been neglected but recently  becomes a serious concern, especially for the alkali-earth atoms. The ground state population loss resulting from the increase of  the pooling  state population is often employed as the experimental observables \cite{RydbergDressingKillian, shannonRydbergloss, PRAChang}.

Without a  theory to  characterize  the dissipative  interacting  system,  in this paper, we provide a  computationally efficient model to describe  not just the short-term but more importantly the long-term dynamics of fully  Rydberg blockaded $\Lambda$-type three-level  atoms  and a  strong probe light field. The model utilizes a decomposition of the coupled dynamical evolution to overcome a high complexity of computing the evolution of the many-body system, leading to convenient calculations and practical analysis in experiments. In  the strong field limit (SFL), we analytically demonstrate the connection between the scattered-field correlation functions and the atomic relaxation,  showing a distinctive quasi-steady-state power spectrum  of the scattered field. For a system without dissipative flipping to unsymmetrical-collective-states, the atomic relaxation time becomes a function of the atom number.  

The paper is organized in the following way. In section \ref{section:singleatom}, we discuss the dynamics of  a  single $\Lambda$-type atom interacting with the probe light field. In section \ref{section:model}, we elaborate the derivation and simplification of the  model which describes the evolution of the fully Rydberg blockaded $\Lambda$-type atoms. In section \ref{section:calculation}, we present the results achieved through the calculations based on the model.  The paper is summarized in section \ref{section:conclusion}. 

\section{A single $\Lambda$-type atom interacting with the light fields} \label{section:singleatom}

\begin{figure}[t]
\centering
     \includegraphics[angle=0,width=0.6\textwidth]{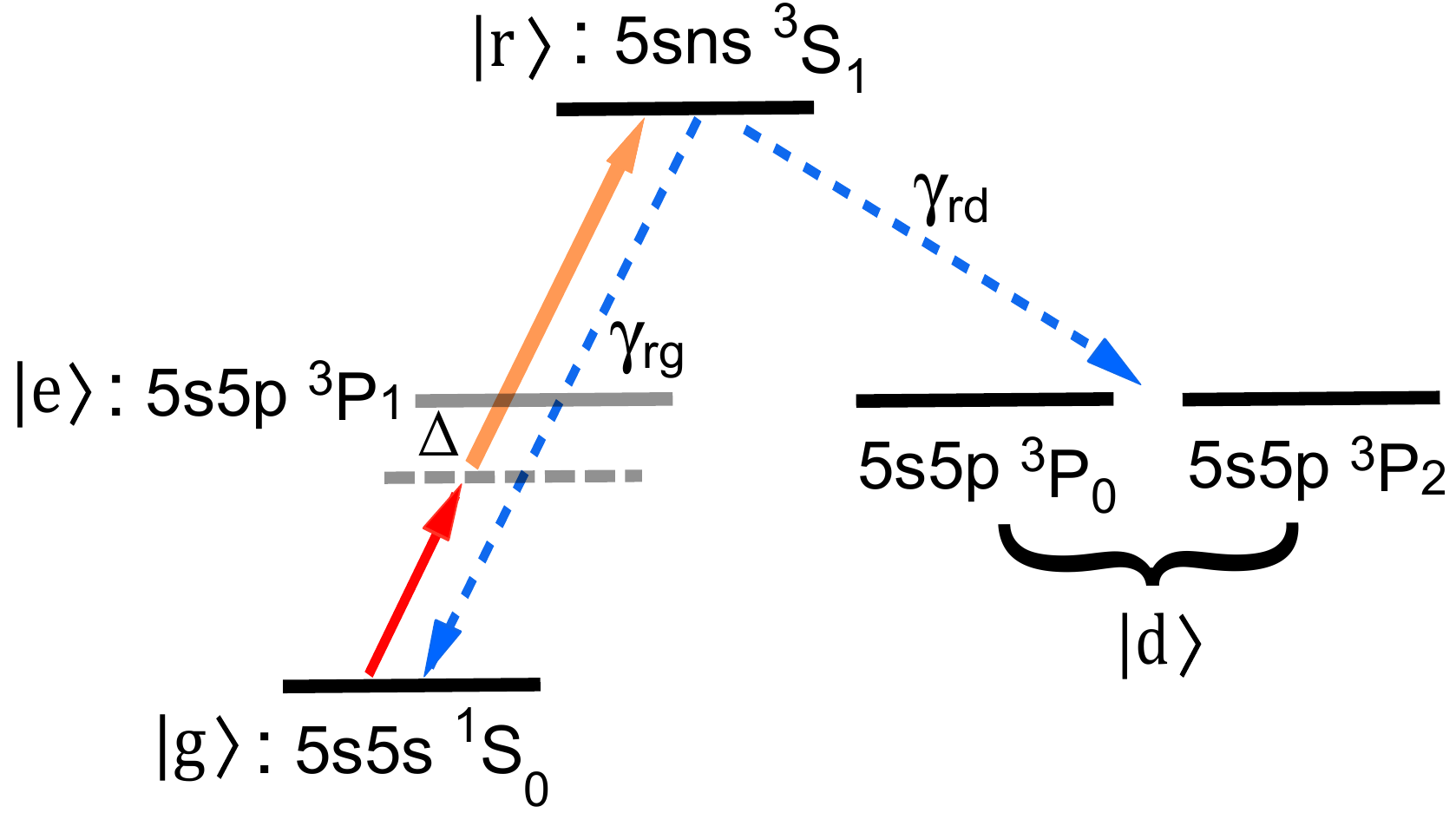}
	\caption{  A typical $\Lambda$-type configuration in the scheme of the Rydberg excitation of the alkaline-earth strontium atom.    }
	\label{fig:sketch0}
\end{figure}
The $\Lambda$-type configuration emerges in the Rydberg excitations of, for example, the alkaline-earth  atoms \cite{RydbergDressingKillian}.   Figure.\ref{fig:sketch0} shows a typical scheme of exciting a strontium atom from the  ground state $|g\rangle$ ($5s5s\,^1 S_0$) to a triplet Rydberg state $|r\rangle$ ($5sns\,^3 S_1$)  by a two-photon Raman process, in which a rather large detuning $\Delta$ leads to an adiabatic elimination of the intermediate state $|e\rangle$ ($5s5p\,^3 P_1$). Due to the coupling of the atom and the environments, the spontaneous decay of the Rydberg state $|r\rangle$  finally transfers the population of $|r\rangle$ to the population of both the ground state $|g\rangle$  and the other two metastable pooling states ($5s5p\,^3 P_0$ and $5s5p\,^3 P_2$). The pooling state $|d\rangle$, decoupled from the light fields,  is represented by all the metastable states, each of which has a lifetime much longer than any of the other time scales of the dynamical evolution. Our discussion starts by considering a single $\Lambda$-type atom interacting with the radiation fields. In order to provide a concise description of the decay from $|r\rangle$ to $|g\rangle$ and the decay from $|r\rangle$ to $|d\rangle$, we neglect any intermediate levels involved in the two decays and use  $\gamma_{rg}$ and  $\gamma_{rd}$ to denote the two overall decay rates respectively.  The  probe field coupling $|g\rangle$ and $|e\rangle$ is in its continuum modes, giving the positive frequency part of the field operator as $ \hat{E}^{(+)}(t) = \int\epsilon_{k}\varepsilon(k) \hat{a}_{k} e^{-i\omega_{k} t}dk$,
where $k$ is the wave vector, $\epsilon_{k}$  the polarization unit vector, $\varepsilon(k)$  the complex amplitude, $\hat{a}_{k}$  the field annihilation operator, $\omega_{k} = c|k|$, and c  the light speed in the vacuum.  It is assumed that a classical pump field coupling $|e\rangle$ and $|r\rangle$ has a Rabi frequency $\Omega_c$.   In the interaction picture and under the rotating-wave approximation, the Hamiltonian governing the time evolution reads 
\begin{eqnarray}
\label{eq:hi}
 \mathscr{H}_{S} = \hbar\int(g_{k} \hat{\sigma}_{rg}\hat{a}_{k}e^{-i(\omega_{k}-\omega_{ge})t} + h.c.)\, dk,
\end{eqnarray}
where  $g_{k} =(\mu_{ge}\cdot \epsilon_{k})\varepsilon(k) \Omega_c/(2\Delta) $ ($\mu_{ge}$ is the atomic dipole moment),  $\hbar\omega_{ab}$ is the energy difference between $|a\rangle$ and $|b\rangle$, and  $\hat{\sigma}_{ab}$=$|a\rangle \langle b |$ hereafter.   Likewise,  the Hamiltonian describing the interaction between the atom and the  environments can be shown in the same way. Considering that  the light field is initially prepared in a continuum coherent state \cite{continuumstate},  by adopting  the Weisskopf-Wigner approximation,  we obtain a set of  Heisenberg equations which describe the time evolution of the operators:
\begin{eqnarray}\label{eq:hsb}
\partial_{t} \hat{\sigma}_{gg}& =  i  \sqrt{ f(t)\gamma}\, (\hat{\sigma}_{rg}-\hat{\sigma}_{gr})+ (\gamma+\gamma_{rg})\hat{\sigma}_{rr},\nonumber\\
\partial_{t} \hat{\sigma}_{rr}& =  i  \sqrt{ f(t)\gamma}\, (\hat{\sigma}_{gr}-\hat{\sigma}_{rg})- (\gamma+\gamma_{rg} + \gamma_{rd}) \hat{\sigma}_{rr},\nonumber\\
\partial_{t} \hat{\sigma}_{gr}& =  i \sqrt{ f(t)\gamma}\, (\hat{\sigma}_{gg}-\hat{\sigma}_{rr})- \frac{\gamma+\gamma_{rg} + \gamma_{rd}}{2}\hat{\sigma}_{gr},
\end{eqnarray}
where $f(t)$ is the input photon rate of the probe field,  $\gamma$ is the rate of photons emitted into the mode of the probe field (i.e.,$\gamma=\int|g_{k}|^2 \delta(k-\omega_{ge}/c)\, dk$) \cite{propagatingQED2}.  Note that $\gamma$ depends on the matching efficiency between  the probe field mode and the atomic emission  mode\cite{PropagatingQED1}.  The terms of the Langevin-type noises are neglected in equation (\ref{eq:hsb}), as the environments are  in the vacuum states.  We assume that $f(t)$ is a step function of time, which leads to a constant Rabi frequency $\Omega=2\sqrt{ f(t)\gamma}$.   Given  the solution of  equation (\ref{eq:hsb}),  the  two-time correlation function $\langle\hat{\sigma}_{rg}(t)\hat{\sigma}_{gr}(t+\tau)\rangle$ can be easily calculated based on the quantum regression theorem. In the SFL, i.e.,  $\Omega\gg\gamma+\gamma_{rg} + \gamma_{rd}$, we introduce $S(\omega)=\gamma\text{Re}[\int_{0}^{+\infty}\langle \hat{\sigma}_{rg}(t)\hat{\sigma}_{gr}(t+\tau) \rangle e^{-i\omega\tau} d\tau]$ to describe a quasi-steady-state power spectrum of the light scattered into the mode of the probe field  \cite{scullybook,QEDSuperatom1}.  With the initial atomic state prepared on $|g\rangle$ and  $\Gamma=\gamma+\gamma_{rg} + \gamma_{rd}$, we find $S(\omega)=e^{-x\gamma_{rd}t}\times A[\Omega,\Gamma]$, where 
\begin{eqnarray}
\label{eq:g1s}
A[\Omega,\Gamma]&=\gamma(\frac{\Gamma/8}{(\omega-\omega_{gr})^2+(\Gamma/2)^2}\nonumber\\
&+\frac{(3\Gamma-\gamma_{rd})/8}{4(\omega-\omega_{gr}-\Omega)^2+(3\Gamma-\gamma_{rd})^2/4}\nonumber\\
&+\frac{(3\Gamma-\gamma_{rd})/8}{4(\omega-\omega_{gr}+\Omega)^2+(3\Gamma-\gamma_{rd})^2/4})
\end{eqnarray}
and $x=\Omega^2/(\Gamma^2+2\Omega^2)$. Equation (\ref{eq:g1s}) is obtained under the condition   $t\gg1/\Gamma$,  i.e., the atom reaches the steady state in the subspace $H_{gr}$ spanned by $|g\rangle$ and $|r\rangle$. As $\Omega> \Gamma > \gamma_{rd}$, $e^{-x\gamma_{rd}t}$  acts as a  slowly-varying amplitude of $S(\omega)$, therefore we refer  $S(\omega)$ as the quasi-steady-state power spectrum.  When $\gamma_{rd}$ is zero,  equation (\ref{eq:g1s}) adapting to a two-level atom  matches perfectly to Mollow's triplet theory \cite{Mollow}. The decay from  $|r\rangle$  to  $|d\rangle$  influences the  linewidths and the amplitudes of the center peak at $\omega = \omega_{gr}$ and the two sidebands at  $\omega = \omega_{gr}\pm\Omega$. The relative heights of the three peaks are altered by  $\gamma_{rd}$ as well. In the weak field limit (WFL), i.e., $\Omega\ll\Gamma$, a single peak appears, i.e., $S(\omega)= e^{-x\gamma_{rd}t}\times\gamma\frac{\gamma_{rd}\Omega^4/\Gamma^4}{(\omega-\omega_{gr})^2+\gamma_{rd}^2\Omega^4/\Gamma^4}$ for  $t\gg1/\Gamma$.

\section{Rydberg blockaded $\Lambda$-type atoms}\label{section:model}

\begin{figure}[t]
\centering
     \includegraphics[angle=0,width=0.65\textwidth]{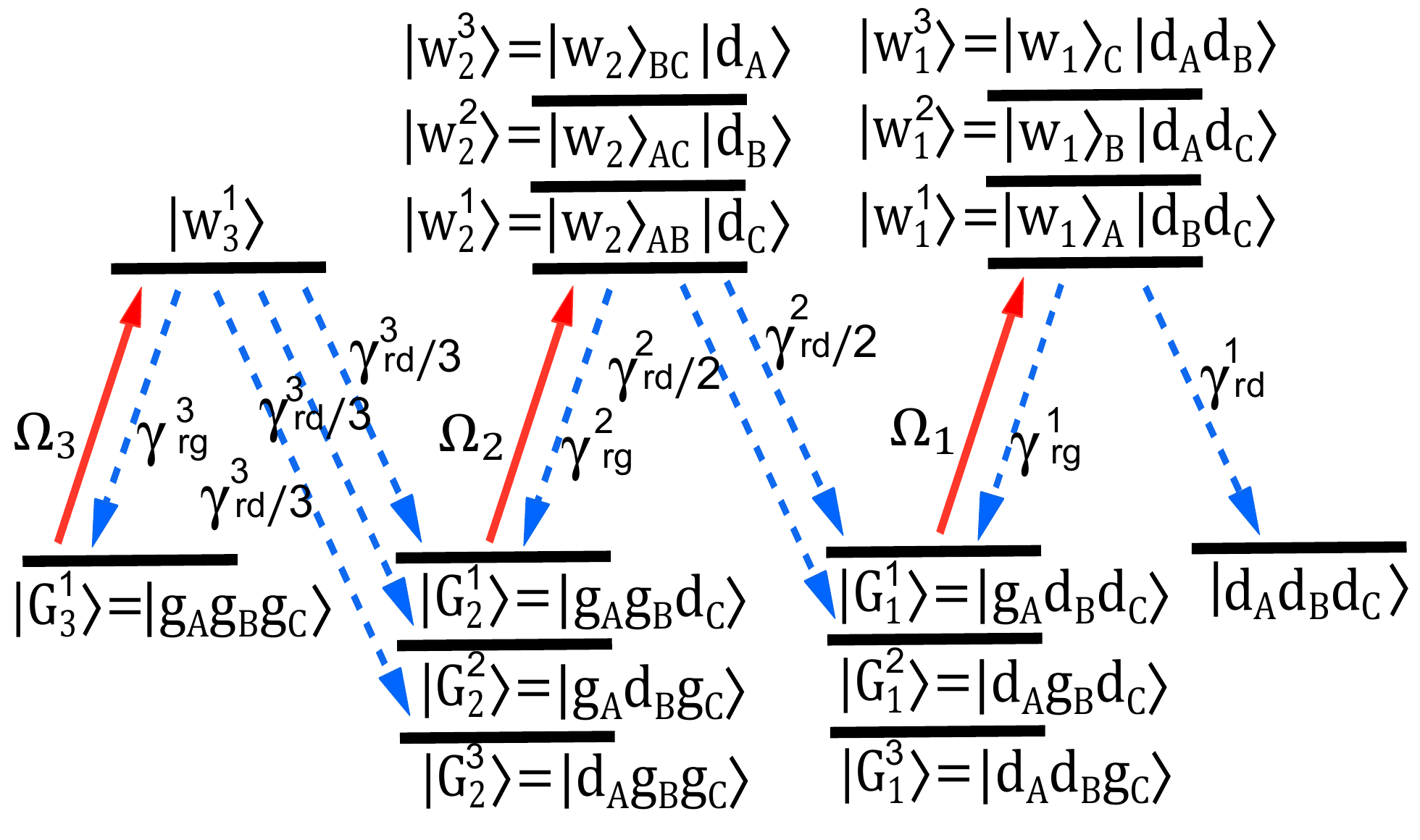}
	\caption{   Truncated level scheme of the fully-Rydberg-blockaded ensemble composed of 3 $\Lambda$-type three-level atoms which are labelled as A,B,and C. Illuminated here is one relaxation path going from the initial state  $|g_Ag_Bg_C\rangle$ to the steady state $|d_Ad_Bd_C\rangle$ through two intermediate states $|g_Ag_Bd_C\rangle$ and $|g_Ad_Bd_C\rangle$ (similar paths going through the other intermediate states are not shown but considered as well). $|W_2\rangle_{xy}=\frac{1}{\sqrt{2}}(|g_x r_y\rangle +|r_x g_y\rangle)$ and $|W_1\rangle_{z}=|r_z\rangle$, where $x$, $y$, $z$ are in $\{A,B,C\}$.  See the text for further elaboration. }
	\label{fig:sketch}
\end{figure}
Based on the above discussion of the single atom case, we extend our study to an atomic ensemble with interatomic Rydberg-Rydberg interactions  which result in the Rydberg blockade effects that only one Rydberg excitation can be accommodated. Assume that each atom couples to the light fields in the same way as in section \ref{section:singleatom}. So a ground state $|g_1,\cdots,g_j\rangle$ with $j$ atoms on the $|g\rangle$ state is only  collectively excited to a symmetrical Dicke state $(1/\sqrt{j})\sum_{m=1}^{j}|g_1,\cdots,r_m,\cdots,g_j\rangle$, where $|r_m\rangle $ is the excited state of the $m$th atom \cite{SuperatomTheory}.  The evolution of an $N$-atom system is depicted by a diagram shown in figure \ref{fig:sketch} (a). We denote $|G^l_j\rangle$ as the $l$th state of all the $N_s$  ground states with $j$ atoms on $|g\rangle$ yet with the other $N-j$ atoms on $|d\rangle$, and denote the only excited state of $|G^l_j\rangle$ as $|W^l_j\rangle$.  Obvious is $N_s[j]=\frac{N!}{(N-j)!j!}$.
In addition, due to dissipations, a $|W^l_j\rangle$ state may incoherently flip to  $j-1$ unsymmetrical states   \cite{collectiveDstate1,collectiveDstate2,collectiveDstate3,QEDSuperatom1}, which then decay to the ground states.   
In the description of  the $N$-body system decay, we apply the same idea as in describing a single atom decay (see figure.\ref{fig:sketch0}), i.e.,  we neglect the dark intermediate states (unsymmetrical states)  and use overall decay rates to describe the decays from the bright states (symmetrical states) to the ground states. As additional decay channels going through the unsymmetrical  states are placed  to  the decay from  $|W^l_j\rangle$ to  $|G^{l^\prime}_{j^\prime}\rangle$ \cite{SuperatomDarkstate1, SuperatomDarkstate2}, the decay from $|W^l_j\rangle$  to  $|G^l_j\rangle$ consists of three parts: the direct decay to the environment,  the decay into the mode of the probe light field,  and indirect decay  through the $j-1$ unsymmetrical states, i.e., $\gamma_{rg}^j=\gamma_{rg}+j\times\gamma+D_{rg}^j$.   And the decay from $|W^l_j\rangle$ to  $|G^{l^\prime}_{j-1}\rangle$  consists of two parts: the direct decay to the environment and indirect decay  through the unsymmetrical states, i.e., $\gamma_{rd}^j/j=\gamma_{rd}/j+D_{rd}^j/j$. All the overall decay rates, $\gamma_{rg}^j$ and $\gamma_{rd}^j$,  are only $j$-dependent but $l$-independent, because the evolution of each single atom is identical.

\begin{figure}[t]
	\centering
     \includegraphics[width=0.6\textwidth]{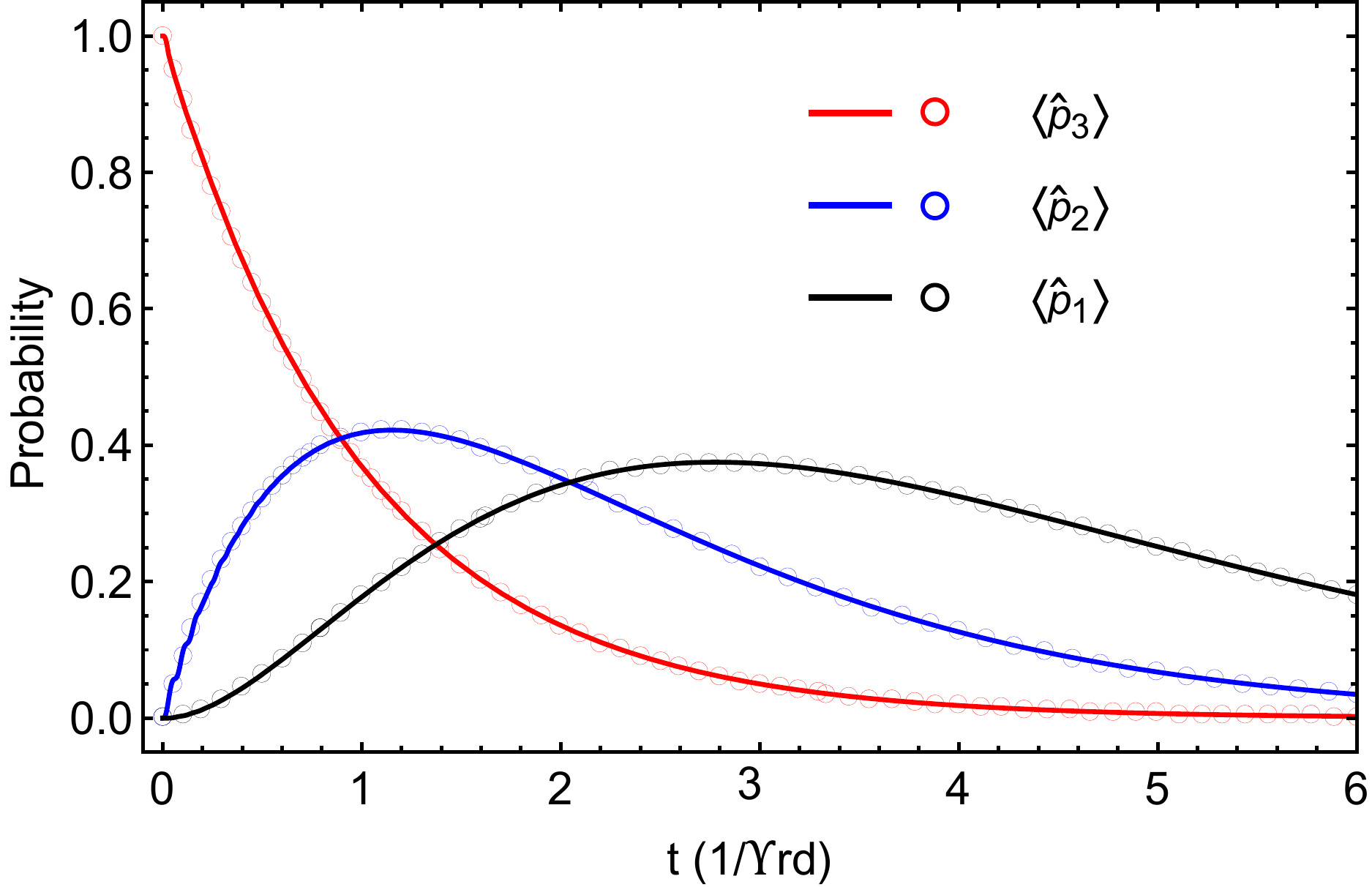}
	\caption{  $\langle\hat{p}_j\rangle $s  as  functions of time with $N=3$, $\Omega=30\gamma_{rd}$, $\gamma=\gamma_{rg}=\gamma_{rd}$ and $D_{rg}^j=D_{rd}^j=(j-1)\gamma_{rd}/2$.    The lines show the numerical solution of equation (\ref{eq:master3}); and the circles represent the results of equation (\ref{eq:pe}).
}
	\label{fig:ta}
\end{figure}

The Hamiltonian  governing the coupling between the  atomic ensemble and the radiation field  reads
\begin{eqnarray}
\label{eq:hi2}
 \mathscr{H}_{N} = \hbar\sum_{j=1}^N\sum_{l=1}^{N_s[j]}\int(\sqrt{j} g_{k} \hat{\sigma}_{W^l_jG^l_j}\hat{a}_{k}e^{-i(\omega_{k}-\omega_{ge})t} + h.c.)\, dk.
\end{eqnarray}
By adopting the same approach leading to Eqs.[\ref{eq:hsb}], we obtain the Heisenberg equations which predicts the time evolution of the system, i.e.,
\begin{eqnarray} \label{eq:master3}
\partial_{t}\tilde{\sigma}_{G_jG_j}&=\frac{i\Omega_j}{2}(\tilde{\sigma}_{G_j W_j}-\tilde{\sigma}_{W_j G_j})
+\gamma^{j+1}_{rd}\tilde{\sigma}_{W_{j+1} W_{j+1}}+\gamma_{rg}^j \tilde{\sigma}_{W_j W_j},\,\,\,\,\,\,\,\,(\text{a})\nonumber\\
\partial_{t}\tilde{\sigma}_{W_jW_j}&=\frac{i\Omega_j}{2}(\tilde{\sigma}_{W_j G_j}-\tilde{\sigma}_{G_j W_j})-\Gamma_j \tilde{\sigma}_{W_j W_j},\,\,\,\,\,\,\,\,\,\,\,\,\,\,\,\,\,\,\,\,\,\,\,\,\,\,\,\,\,\,\,\,\,\,\,\,\,\,\,\,\,\,\,\,\,\,\,\,\,\,\,\,\,(b)\nonumber\\
\partial_{t}\tilde{\sigma}_{G_jW_j}&=\frac{i\Omega_j}{2}(\tilde{\sigma}_{G_j G_j}-\tilde{\sigma}_{W_j W_j})-\frac{\Gamma_j}{2}\tilde{\sigma}_{G_jW_j},\,\,\,\,\,\,\,\,\,\,\,\,\,\,\,\,\,\,\,\,\,\,\,\,\,\,\,\,\,\,\,\,\,\,\,\,\,\,\,\,\,\,\,\,\,\,\,\,\,\,\,\,\,(c)
\end{eqnarray}
where   $\tilde{\sigma}_{\alpha_j\beta_j}=\sum_{l=1}^{N_s[j]}\hat{\sigma}_{\alpha^l_j\beta^l_j}$ ($\alpha,\beta\in\{G,W\}$), $\Omega_j=\sqrt{j}\Omega$,  $\Gamma_j=\gamma_{rg}^j+\gamma_{rd}^j$,  $j\in\{1,\cdots,N\}$ and $ \tilde{\sigma}_{W_{N+1} W_{N+1}}= 0$  (see \ref{appendixa}). It is easy to see $\langle \tilde{\sigma}_{\alpha_j \beta_j}\rangle=N_s[j]\times\langle\hat{\sigma}_{\alpha^k_j \beta^k_j}\rangle\,\forall\, k\in\{1,\cdots,N_s[j]\}$. 
The time evolution of the system shows an atomic relaxation transferring the initial state $|g,\cdots,g\rangle$ to the steady state $|d,\cdots,d\rangle$.  The following discussions are  mainly based on the SFL  ($\Omega_j\gg \Gamma_j\,\forall\, j\in\{1,\cdots,N\}$)  in which a more efficient relaxation occurs.  In a realistic situation of exciting the ground state strontium atoms to a  Rydberg state $5s30s ^3S_1$  ($\gamma_{rg}\approx 2\pi\times5$ kHz, $\gamma_{rd}\approx 2\pi\times10$ kHz), it is likely to have $\Omega=2\pi\times 500$  kHz and $N\sim \sqrt{C_6/\hbar\Omega }\times\rho_d\sim30$, where $C_6/\hbar=2\pi\times 20$ MHz$\cdot\mu m^6$ and  the atomic density $\rho_d=5\times 10^{12}/cm^3$  \cite{PRAChang}.   In this system,  based on the estimation $\Gamma_j\sim j(2\gamma_{rg}+\gamma_{rd})$, where $D_{rg}^j \sim (j-1)\gamma_{rg}$ and $D_{rd}^j \sim (j-1)\gamma_{rd}$), it can be inferred that the SFL is well satisfied ($\sqrt{j}\Omega\gg\Gamma_j $).  We define an operator  $\hat{p}_j=\tilde{\sigma}_{G_jG_j}+\tilde{\sigma}_{W_jW_j}$, the evolution of which gives a direct reflection of the atomic relaxation. From equation (\ref{eq:master3})  (a) and (b),  $\partial_{t}\,\hat{p}_j=-\gamma^j_{rd}\tilde{\sigma}_{W_jW_j}+\gamma^{j+1}_{rd}\tilde{\sigma}_{W_{j+1}W_{j+1}}$.  In the SFL, as  $\langle\tilde{\sigma}_{W_jW_j}\rangle = \langle\hat{p}_j\rangle\,\Omega_j^2/(\Gamma_j^2+2\Omega_j^2)+\delta_j$, where $\delta_j$ denotes a rapidly-varying term resulting from the Rabi oscillation,  we extract the equations of motion about  $\langle \hat{p}_j\rangle$ from equation (\ref{eq:master3})  by adiabatically eliminating  $\delta_{j}$ and by using $\Omega_j^2/(\Gamma_j^2+2\Omega_j^2)=1/2$:    
\begin{eqnarray}
\label{eq:pe}
\partial_t\, \langle\hat{p}_j\rangle= -(\gamma_{rd}^j\langle\hat{p}_j\rangle-\gamma_{rd}^{j+1}\langle\hat{p}_{j+1}\rangle)/2.
\end{eqnarray}
With the initial condition $\langle\hat{p}_{N}(0)\rangle=1$ and $\langle\hat{p}_{j}(0)\rangle|_{j\neq N}=0$, $\langle\hat{p}_j\rangle$s are  fully determined  by  equation (\ref{eq:pe}) (see figure \ref{fig:ta}). For a system with negligible dissipative flipping to unsymmetrical-collective-states, i.e.,  $\gamma_{rd}^j=\gamma_{rd}\,\forall j\in\{1,\cdots,N\} $,  solving equation (\ref{eq:pe})  yields
\begin{eqnarray}
\label{eq:ps}
\langle\hat{p}_j\rangle = \frac{(\gamma_{rd}\,t)^{N-j}}{2^{N-j}\times(N-j)!}\times e^{-\gamma_{rd}\,t/2}.
\end{eqnarray}
After replacing equation (\ref{eq:master3})  (a) with $\tilde{\sigma}_{G_jG_j}+\tilde{\sigma}_{W_jW_j}=\hat{p}_j$ and extracting the average value of $\hat{p}_j$ from equation (\ref{eq:pe}), we  thus find a \emph{\textbf{decomposition}} of the dynamical evolution  in equation (\ref{eq:master3}): the interaction between the light field and the connected multi-level atomic system is decomposed into interactions between the light field and  open two-level systems, each of which evolves in a discrete subspace $H_{jl}$ spanned by  $|G^l_j\rangle$ and $|W^l_j\rangle$.
The decomposition provides  convenience to calculate the scattered field correlation functions and the atomic state fractions in section \ref{section:calculation}. 

\section{Power spectrum and atomic fractions}\label{section:calculation}
\begin{figure}[t]
	\centering
	\includegraphics[width=0.6\textwidth]{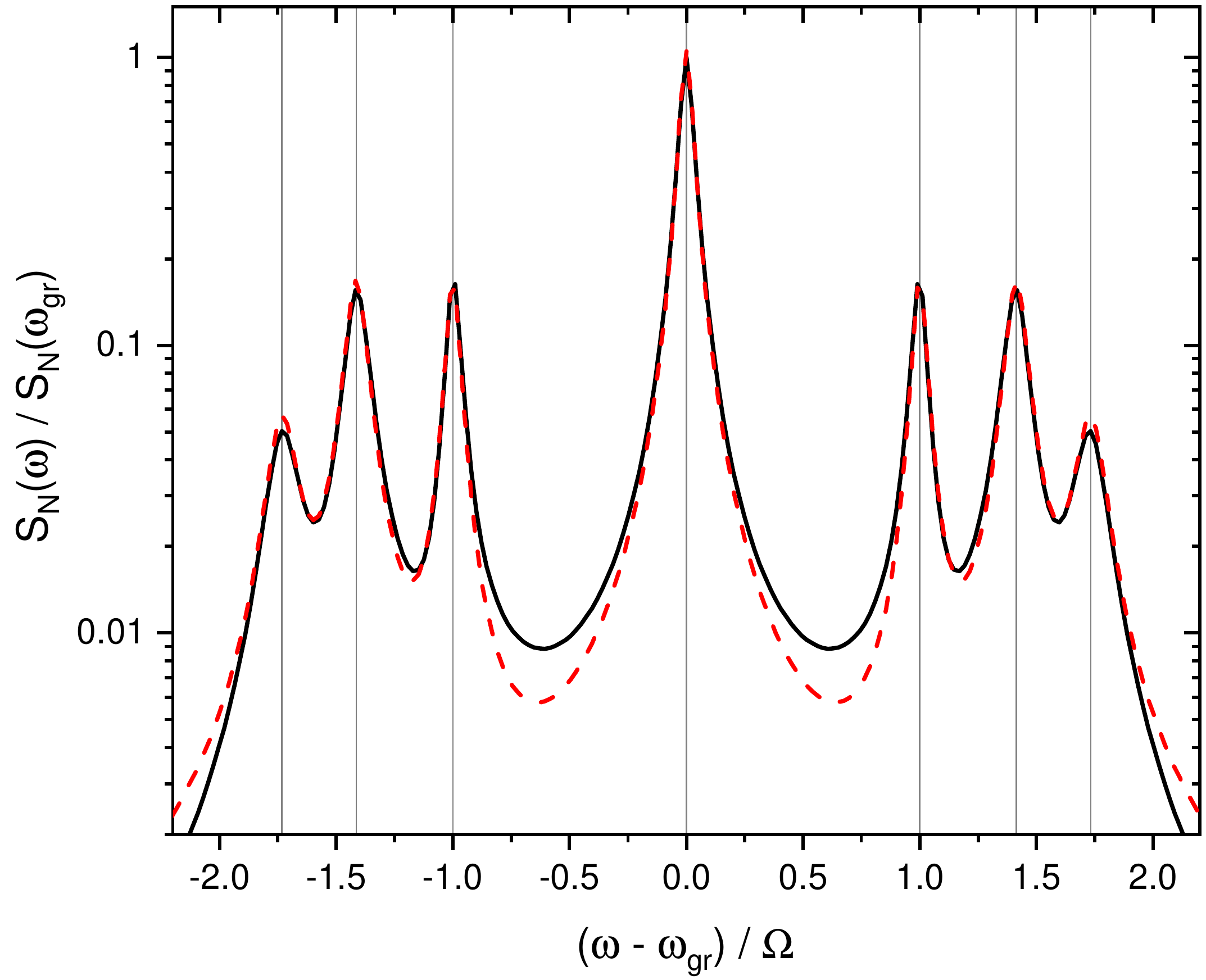}
	\caption{ Normalized $S_{N}(\omega)$ with $N=3$. The normalization coefficient $S_{N}(\omega_{gr})$ is  given numerically by the solution of equation (\ref{eq:master3}).   The results obtained  numerically by equation (\ref{eq:master3}) and  analytically by  equation (\ref{eq:ftwo}) are shown by the solid black curve and the red dashed curve respectively.   The vertical lines correspond to $\omega - \omega_{gr} =\pm \sqrt{j}\, \Omega$ with $ j\in \{0,1,2,3\}$.  In the calculation, $\Omega=30\gamma_{rd}$, $\gamma=\gamma_{rg}=\gamma_{rd}$, $D^j_{rg}=D^j_{rd}=(j-1)\gamma_{rd}/2\, \forall j\in\{1,2, 3\}$, and $t=5/(\gamma_{rg}+\gamma_{rd})$.  }
	\label{fig:es}
\end{figure}

We  examine the correlation functions of the scattered field. Under the Weisskopf-Wigner approximation,  the operator indicating the positive frequency part of the field scattered into the probe field mode is 
\begin{eqnarray}
\label{eq:es}
\hat{E}_s^{(+)}(t)=  -i\sum_{j=1}^N\sqrt{j\gamma}\,\tilde{\sigma}_{G_jW_j}.
\end{eqnarray}
In the SFL, due to the decomposition,  we can calculate the $N$ operators  ($\tilde{\sigma}_{G_NW_N},\cdots, \tilde{\sigma}_{G_jW_j}, \cdots, \tilde{\sigma}_{G_1W_1}$) separately, which in addition implies that each of the $N_s[j]$ operators ($\hat{\sigma}_{G^1_jW^1_j},\cdots, \hat{\sigma}_{G^l_jW^l_j},\cdots,\hat{\sigma}_{G^{N_s[j]}_jW^{N_s[j]}_j}$)  evolves in its own  subspace $H_{jl}$ in the same way.
As  $\Omega_j\gg\gamma_{rg}^j\gg\gamma_{rd}^j\, \forall\,j\in\{1,\cdots,N\}$, the evolution of $\hat{p}_j$ goes more slowly than that of $\tilde{\sigma}_{G_{j}W_{j}}$. Hence the atomic system reaches a quasi-steady state for $t\gg 1/\Gamma_j\, \forall\,j\in\{1,\cdots,N\}$, which indicates a steady state in each $H_{jl}$.  In this sense, we find the quasi-steady-state power spectrum  $S_{N}(\omega)$ of the field scattered by the $N$-body interacting system, i.e.,  $S_{N}(\omega)=\text{Re}[\int_{0}^{\infty}\langle \hat{E}_s^{(-)}(t)\hat{E}_s^{(+)}(t+\tau)\rangle e^{-i\omega\tau}d\tau] $ where $t\gg 1/\Gamma_j\, \forall\,j\in\{1,\cdots,N\}$. Due to  $\langle\tilde{\sigma}_{G_{j_1}W_{j_1}}\tilde{\sigma}_{W_{j_2}G_{j_2}}\rangle|_{j_1\not=j_2}\approx0$,         
\begin{eqnarray}
\label{eq:ftwo}
S_{N}(\omega)& \approx \gamma\sum_{j=1}^N j\,\text{Re}\,[\int_{0}^{\infty}\langle\tilde{\sigma}_{G_{j}W_{j}}(t)\tilde{\sigma}_{W_{j}G_{j}}(t+\tau)\rangle e^{-i\omega\tau}d\tau] \nonumber\\
&\approx\sum_{j=1}^N \, j\,\langle\hat{p}_j\rangle\,A[\Omega_j,\Gamma_j],
\end{eqnarray}
where $\langle\hat{p}_j\rangle$ is given by equation (\ref{eq:pe}) and  $A[\Omega_j,\Gamma_j]$ is found in  equation (\ref{eq:g1s}).     In figure \ref{fig:es},  $S_{N}(\omega)$ with $2N$ sidebands and one central peak is obtained analytically from equation (\ref{eq:ftwo}), which is in good agreement with the numerical solution of equation (\ref{eq:master3}). We see that $S_N(\omega)$ reflects a  partial reconstruction  of the atomic density matrix  $\rho$ in real time since $\langle\hat{p}_j\rangle=\sum_{l=1}^{N_s[j]}\text{Tr} [\rho |G^l_j\rangle\langle G^l_j|+\rho|W^l_j\rangle\langle W^l_j|]$, which analytically reveals the connection between the scattered field quasi-steady-state power spectrum and the atomic relaxation. The $j$th two sidebands, which have the peaks centered at $\omega=\omega_{gr}\pm\sqrt{j}\Omega$ and  the width of about $3\Gamma_j/2$,   exhibit the same slowing-varying amplitudes determined by $\langle\hat{p}_j\rangle$.  As a feature found neither in a two-level superatom nor in a single $\Lambda$-type atom,  the relative heights of the  $j$th two sidebands and the $j^{\prime}$th two sidebands  change significantly over time due to the relative change of $\langle\hat{p}_j\rangle$ and  $\langle\hat{p}_{j^{\prime}}\rangle$  (see figure \ref{fig:ta}).  This observable information carried by the scattered field reflects the atomic evolution time $t$ and implies special applications of embedding the interacting $\Lambda$-type system in  quantum networks \cite{QuantumNetworks, LukinRMP}. The application also can be demonstrated by the measurements of higher-order scattered field correlation functions,  which show a time-dependence determined by $\langle\hat{p}_j\rangle$s as well. In addition,  we consider that the Rydberg excitation fields are switched off in the middle.   After the  spontaneous decays of the Rydberg states, the atomic density matrix  becomes  $\rho\approx \sum_{j=0}^{N}\sum_{l=1}^{N_s[j]}\frac{\langle\hat{p}_j\rangle}{N_s[j]} |G^l_j\rangle\langle G^l_j|$, where $\langle\hat{p}_0\rangle=1- \sum_{j=1}^{N} \langle\hat{p}_j\rangle$. Thus, the fields turned on again at a time $t^{\prime}$ drive $N$ independent Rabi oscillations,  each of which goes in its own Hilbert space with a Rabi frequency $\sqrt{j} \Omega$. The Rydberg fraction is given as  $P_r(t^{\prime}+\Delta t^{\prime})\approx\frac{1}{2N}\sum_{j=1}^{N}\langle\hat{p}_j\rangle(1 - \text{Cos}[\sqrt{j} \Omega \Delta t^{\prime}])$ for $\Delta t^{\prime}\ll1/\Gamma_j\forall j\in \{1,\cdots,N\}$, showing another way to determine $\langle\hat{p}_j\rangle$ by  $P_r(t^{\prime}+\Delta t^{\prime})$. As a point   linked to the multiple Rabi oscillations,  the second-order correlation function which can be written as a  superposition based on the decomposition,  i.e., $\langle \hat{E}_s^{(-)}(t^{\prime}_1)\hat{E}_s^{(-)}(t^{\prime}_2)\hat{E}_s^{(+)}(t^{\prime}_2)\hat{E}_s^{(+)}(t^{\prime}_1)\rangle 
\approx \sum_{j=1}^N \gamma^2 j^2 \langle \tilde{\sigma}_{G_{j}W_{j}}(t^{\prime}_1)\tilde{\sigma}_{G_{j}W_{j}}(t^{\prime}_2)\tilde{\sigma}_{W_{j}G_{j}}(t^{\prime}_2)\tilde{\sigma}_{W_{j}G_{j}}(t^{\prime}_1)\rangle $, exhibits different periodic alternations of photon bunching and antibunching as compared with that of a  2 dimensional superatom \cite{QEDSuperatom1,LukinRMP},   demonstrating  distinctive  photon-photon correlations generated by the Rydberg blockaded $\Lambda$-type atoms.

As what can be observed from the experiments of exciting strongly Rydberg blockaded $\Lambda$-type atoms\cite{PRAChang},    the increasing rate of the pooling state fraction remains almost constant when it increases from zero to more than one half.   Based on our modelling, a special system with negligible  dissipative flipping to unsymmetrical-collective-states has a similar performance. With $D^j_{rg}=D^j_{rd}=0$, the fractions of the atoms in different atomic states can be deduced by equation (\ref{eq:ps}).  The $|d\rangle$ state  fraction  is  $P_d(t)=1-\sum_{j=1}^N\frac{j}{N}\langle\hat{p}_j\rangle$. The population on the steady state $|G_0\rangle=|d,\cdots,d\rangle$ is $P_{G0}(t)=1-\sum_{j=1}^{N}\langle\hat{p}_j\rangle$.  The time derivative of $P_d(t)$ is thus written as 
\begin{eqnarray}
\label{eq:dp}
\frac{d\,P_d(t)}{dt}= \frac{\gamma_{rd}}{2N}(1 - P_{G0}(t)). 
\end{eqnarray}
Note that the fraction of the Rydberg atoms is  $P_r(t)=\frac{1}{N}\sum_{j=1}^{N}\langle\tilde{\sigma}_{W_jW_j}\rangle$, i.e., $P_r(t)\approx\sum_{j=1}^{N}\frac{\langle\hat{p}_j\rangle}{2N} - \frac{\langle\hat{p}_N\rangle}{2N}\text{Cos}[\sqrt{N} \Omega t]\,\text{Exp}[-3\gamma_{rg}^{N}t/4]$, where the second term results from the damped Rabi oscillation between $|G^1_N\rangle$ and  $|W^1_N\rangle$. By  eliminating the second rapidly-varying term in $P_r(t)$,  equation (\ref{eq:dp}) can be equivalently described by $\frac{d\,P_d(t)}{dt}=\gamma_{rd} P_r(t)$. In equation (\ref{eq:dp}),  $d P_d(t)/dt$  serves as a prominent indication of the relaxation time $t_r$, at which the population of the steady state $|G_0\rangle$ starts to be non-negligible.    With $t<t_r$, namely $P_{G0}(t)\ll1$,  $P_d(t)$ increases linearly over time with a constant increasing rate $\gamma_{rd}/(2N)$,  i.e., $P_d(t) = \gamma_{rd} t/(2N)$.  However, with $t>t_r$,  the state $|G_0\rangle$ becomes significantly populated,  which brings the evolution of $P_d(t)$ into a non-linear regime.  The relaxation time $t_r$ can be analytically calculated as a function of the atom number.   By using Stirling's approximation  (see    \ref{appendixb}),  we manage to quantitatively find  $t_r$   as  
\begin{eqnarray}
\label{eq:tr}
t_{r}=\frac{2N}{\gamma_{rd}}\sqrt[2N]{2\pi N}(1-\frac{2}{\sqrt{N}}).
\end{eqnarray} 
For $1/\gamma_{rg}\ll t<t_r$,   the Rydberg fraction  remains unperturbed by the atomic relaxation (i.e., $P_r(t) =  \frac{1}{2N}$),  which means the increase of atoms on $|d\rangle$  comes  entirely  from the decrease of atoms on $|g\rangle$. 
 \begin{figure}[t]
	\centering
     \includegraphics[width=0.6\textwidth]{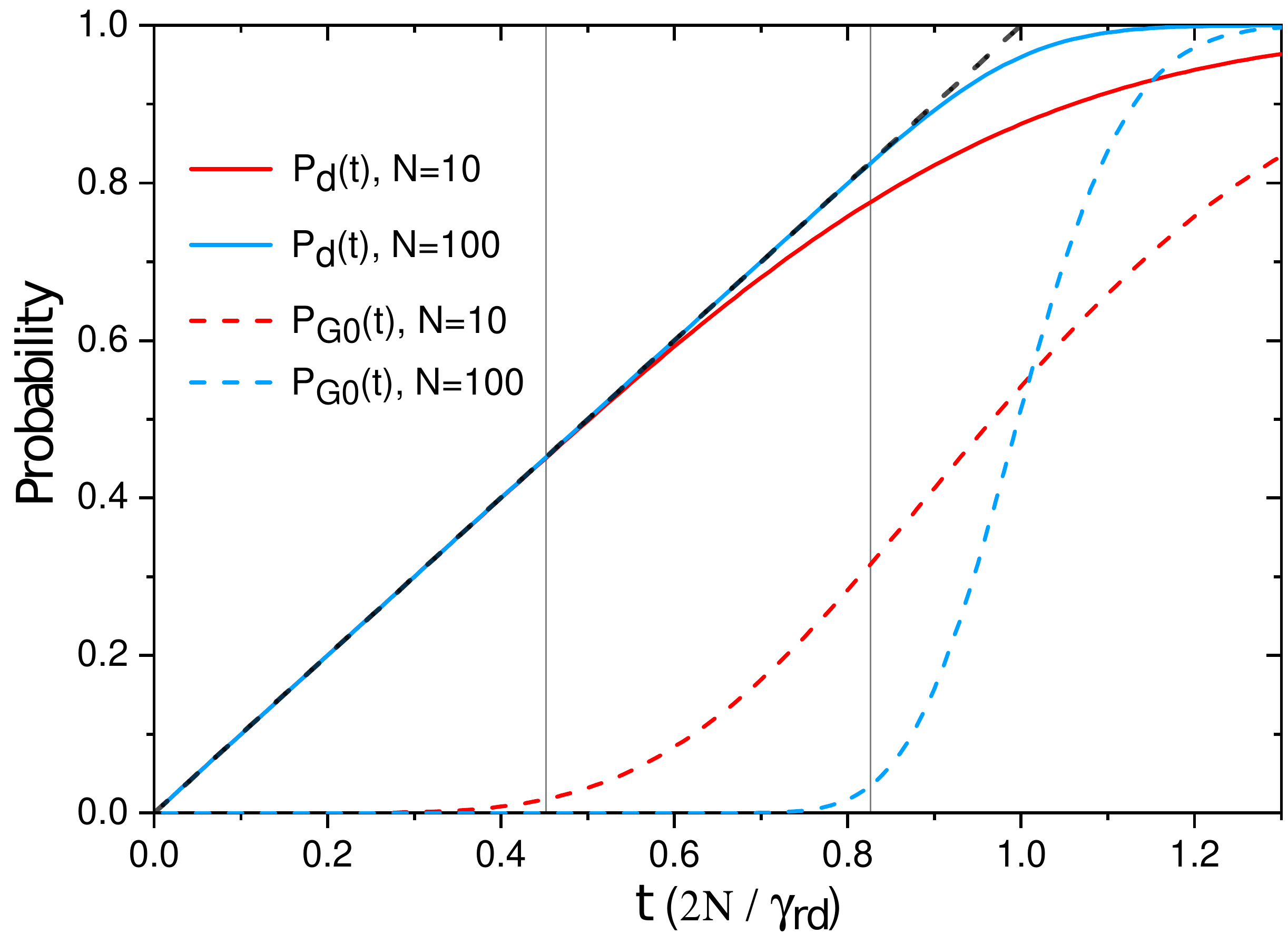}
	\caption{ $P_{d}(t)$ and $P_{G0}(t)$ as  functions of time with  $D_{rg}^j=D_{rd}^j=0\,\forall j\in\{1,\cdots,N\} $.  The linear relation of $P_{d}(t)=\gamma_{rd} t/2N$ is shown by the black dashed line. Two solid vertical lines indicate the relaxation times in equation (\ref{eq:tr}) with $N=10$ and $N=100$.       
}
	\label{fig:ls}
\end{figure}
In figure \ref{fig:ls},  the time evolutions of $P_d(t)$ are shown with different atom numbers. The regime where $P_d(t)$ performs as a linear function of time is well-defined by $t\in(0, t_r)$.  When $N$ is larger, the value of $P_d(t_{r})$ goes closer to $1$. In other words,  the linear relation $P_d(t) = \gamma_{rd} t/(2N)$ becomes a better approximation  to the full extent of the dynamical evolution.  In the non-linear regime ($t>t_r$), the  rate $d P_d(t)/dt$ decreases from  the value $ \gamma_{rd}/(2N)$,  due to the increase of $P_{G0}$. In the experiments regarding Rydberg-dressed $\Lambda$-type atoms \cite{RydbergDressingKillian}, while the total Rydberg state decay rate is previously given by the Rydberg-state-lifetime measurements \cite{Rydberglifetime},   $\gamma_{rd}$  becomes an uncertain but essential variable participating in the experimental simulation.  As $N\gg1$  can be practically fulfilled  by a high enough atomic density  though the SFL  restricts the blockade volume \cite{SuperatomEvidence,QEDSuperatom1},   the  linear regime of $P_d(t)$ indicates a dependable and effective way to experimentally  determine $\gamma_{rd}$. For non-negligible dissipative flipping to unsymmetrical-collective-states, there exists $\gamma_{rd}^j>\gamma_{rd}^{j^{\prime}}>\gamma_{rd}$ with $j>j^{\prime}>1$, which  breaks $P_d(t) = \gamma_{rd} t/(2N)$ and shortens  the relaxation time in Eqs.[\ref{eq:tr}].

\section{Conclusion}\label{section:conclusion}
To sum up,  we provide a model about the dynamics of   fully Rydberg blockaded $\Lambda$-type atoms and a  strong probe light field in a coherent state. In the modelling,  a decomposition of the coupled dynamical evolution unveils analytically a connection between the scattered field correlation functions and the atomic relaxation, which reduces a high complexity of computing the multi-level system evolution. A quasi-steady-state power spectrum of the scattered field are found  with  multiple sidebands, the relative heights of which show a time-dependence  and illuminate different applications in information transfer of quantum networks.  In a special case that  the dissipative flipping to unsymmetrical-collective-states is negligible, the atomic relaxation time  in the SFL becomes a function of the atom number.  

By now, we focus on the dynamical evolution of the $\Lambda$-type atoms fully confined in the Rydberg blockade volume, within which no more than one Rydberg excitation is accommodated. We look forward to further investigations about the multiple Rydberg excitations of an ensemble distributed in a much larger volume, which consists of new subjects in the handling of the long-range interactions between Rydberg atoms \cite{EITsuperatom}. We demonstrate  $N$ independent Rabi oscillations, which may illuminate distinctive photon-photon correlations and bring particular performances of photon-state-manipulation in mesoscopic level \cite{LukinRMP}. In the WFL, the $\Lambda$-type atoms on Rydberg-dressed ground states \cite{RDressingTPfau,RydbergDressingJohnson} acting as a dipolar dissipative system also present a next step of studying the fascinating many-body dynamics in equilibrium.  

We thank Fen Wu, Matthias Weidem$\ddot{\text{u}}$ller, Xingzhong Zhao, Shengshi Pang, Xibo Zhang  for discussions. C.Q. is supported by the National Key Research and Development Program of China under Grant No. 2018YFA0305601. W.Z. is supported by the NSFC (Grants Nos. 91836101 and U1930201).

\begin{appendix}

\section{ Derivation of equation (\ref{eq:master3})}\label{appendixa}
Combining the Heisenberg equations for the time derivatives of the $N_s[j]$ operators ($\hat{\sigma}_{G^1_jG^1_j},\cdots, \hat{\sigma}_{G^l_jG^l_j},\cdots,\hat{\sigma}_{G^{N_s[j]}_jG^{N_s[j]}_j}$) yields
\begin{eqnarray}\label{eq:master2}
\partial_{t}\sum_{l=1}^{N_s[j]}\hat{\sigma}_{G^l_jG^l_j}&=\frac{i\Omega_j}{2}\sum_{l=1}^{N_s[j]}(\hat{\sigma}_{G^l_j W^l_j}-\hat{\sigma}_{W^l_j G^l_j})\nonumber\\
&+\frac{\gamma^{j+1}_{rd}}{j+1}\sum_{l^{\prime}=1}^{N_s[j+1]} N_d[j+1]\hat{\sigma}_{W^{l^{\prime}}_{j+1} W^{l^{\prime}}_{j+1}}+\gamma_{rg}^j \sum_{l=1}^{N_s[j]}\hat{\sigma}_{W^l_j W^l_j}.
\end{eqnarray}
In equation (\ref{eq:master2}),  $N_d[j+1]$ is given by the number of the decays  from one $|W^{l^{\prime}}_{j+1}\rangle$ state to the  states in $\{|G^{1}_{j}\rangle,\cdots,|G^{l}_{j}\rangle,\cdots,|G^{N_s[j]}_{j}\rangle\}$, i.e.,  $N_d[j+1]=j+1$. With the definition of $\tilde{\sigma}_{\alpha_j\beta_j}=\sum_{l=1}^{N_s[j]}\hat{\sigma}_{\alpha^l_j\beta^l_j}$ ($\alpha,\beta\in\{G,W\}$), equation (\ref{eq:master2}) is thus expressed as  equation (\ref{eq:master3}) (a). 

\section{ Derivation of equation (\ref{eq:tr})}\label{appendixb}
Because $P_{d}(t)\le1$,   $t_r<\frac{2N}{\gamma_{rd}}$ can be deduced by equation (\ref{eq:dp}). We write down
$P_{G0}(t)$  as a summation, i.e.,  $P_{G0}(t)= \sum\limits_{j=N}^{+\infty}f[j,t]$, where $f[j,t]=\frac{(\gamma_{rd}\,t/2)^{j}}{j!} e^{-\gamma_{rd}\,t/2}$.  Compared with the other terms in the summation,   $f[N,t]$  diverges first from $0$ when t increases from $0$ to $\frac{2N}{\gamma_{rd}}$.  We define $t_c$ by the solution of $\frac{(\gamma_{rd}\,t/2)^{N}}{N!}=1$, which gives $t_c\approx \frac{2N}{e\gamma_{rd}}\sqrt[2N]{2\pi N}$  by Stirling's approximation.  Replacing $t$ in $f[N,t]$ by $\eta\, t_c$ yields a concise expression $f[N,\eta] = \eta^{N}\,e^{-N\eta/e}$. We find that $f[N,\eta]$ increases monotonically from $0$ to the maximum value $1$ when $\eta$ increases from $0$ to $e$. In addition, the derivative $d f[N,\eta]/d\eta$  reaches the maximum value at $\eta = e-\frac{e}{\sqrt{N}}$.  Therefore, the divergence of $f[N,\eta]$ starts around $\eta = e-\frac{2e}{\sqrt{N}}$, which gives $t_r=(e-\frac{2e}{\sqrt{N}})t_c$ (i.e.,  equation (\ref{eq:tr})). 
\end{appendix}

\section*{References}
\bibliographystyle{unsrt}
\bibliography{iopart-num}

\providecommand{\newblock}{}
\begin{thebibliography}{10}
\expandafter\ifx\csname url\endcsname\relax
  \def\url#1{{\tt #1}}\fi
\expandafter\ifx\csname urlprefix\endcsname\relax\def\urlprefix{URL }\fi
\providecommand{\eprint}[2][]{\url{#2}}

\bibitem{Dicke}
Dicke R~H 1954 {\em Phys. Rev.\/} {\bf 93}(1) 99--110

\bibitem{Scully}
Svidzinsky A~A, Chang J~T and Scully M~O 2008 {\em Phys. Rev. Lett.\/} {\bf
  100}(16) 160504

\bibitem{SuperatomTheory}
Lukin M~D, Fleischhauer M, Cote R, Duan L~M, Jaksch D, Cirac J~I and Zoller P
  2001 {\em Phys. Rev. Lett.\/} {\bf 87}(3) 037901

\bibitem{SuperatomEvidence}
Heidemann R, Raitzsch U, Bendkowsky V, Butscher B, L\"ow R, Santos L and Pfau T
  2007 {\em Phys. Rev. Lett.\/} {\bf 99}(16) 163601

\bibitem{propagatingQED2}
Domokos P, Horak P and Ritsch H 2002 {\em Phys. Rev. A\/} {\bf 65}(3) 033832

\bibitem{QEDSuperatom1}
Paris-Mandoki A, Braun C, Kumlin J, Tresp C, Mirgorodskiy I, Christaller F,
  B\"uchler H~P and Hofferberth S 2017 {\em Phys. Rev. X\/} {\bf 7}(4) 041010

\bibitem{QEDSuperatom2}
Stiesdal N, Kumlin J, Kleinbeck K, Lunt P, Braun C, Paris-Mandoki A, Tresp C,
  B\"uchler H~P and Hofferberth S 2018 {\em Phys. Rev. Lett.\/} {\bf 121}(10)
  103601

\bibitem{singlephotonemitterLukin}
Gorshkov A~V, Otterbach J, Fleischhauer M, Pohl T and Lukin M~D 2011 {\em Phys.
  Rev. Lett.\/} {\bf 107}(13) 133602

\bibitem{QuantumNetworks}
Reiserer A and Rempe G 2015 {\em Rev. Mod. Phys.\/} {\bf 87}(4) 1379--1418

\bibitem{LukinRMP}
Lukin M~D 2003 {\em Rev. Mod. Phys.\/} {\bf 75}(2) 457--472

\bibitem{photonicDissipative1}
Gorshkov A~V, Nath R and Pohl T 2013 {\em Phys. Rev. Lett.\/} {\bf 110}(15)
  153601

\bibitem{photonicDissipative2}
Zeuthen E, Gullans M~J, Maghrebi M~F and Gorshkov A~V 2017 {\em Phys. Rev.
  Lett.\/} {\bf 119}(4) 043602

\bibitem{RydbergDressing2}
Pupillo G, Micheli A, Boninsegni M, Lesanovsky I and Zoller P 2010 {\em Phys.
  Rev. Lett.\/} {\bf 104}(22) 223002

\bibitem{RydbergDressing3}
Henkel N, Nath R and Pohl T 2010 {\em Phys. Rev. Lett.\/} {\bf 104}(19) 195302

\bibitem{RydbergDressingMOTMPA}
Bounds A~D, Jackson N~C, Hanley R~K, Faoro R, Bridge E~M, Huillery P and Jones
  M~P~A 2018 {\em Phys. Rev. Lett.\/} {\bf 120}(18) 183401

\bibitem{RDressingTPfau}
Balewski J~B, Krupp A~T, Gaj A, Hofferberth S, Löw R and Pfau T 2014 {\em New
  Journal of Physics\/} {\bf 16} 063012

\bibitem{RydbergDressing1}
Honer J, Weimer H, Pfau T and B\"uchler H~P 2010 {\em Phys. Rev. Lett.\/} {\bf
  105}(16) 160404

\bibitem{RydbergDressingKillian}
Gaul C, DeSalvo B~J, Aman J~A, Dunning F~B, Killian T~C and Pohl T 2016 {\em
  Phys. Rev. Lett.\/} {\bf 116}(24) 243001

\bibitem{scullybook}
Scully M~O and Zubairy I~H 1997 {\em {Quantum Optics}\/} vol 104 (Cambridge
  University Press, Cambridge)

\bibitem{shannonRydbergloss}
Arias A, Lochead G, Wintermantel T~M, Helmrich S and Whitlock S 2019 {\em Phys.
  Rev. Lett.\/} {\bf 122}(5) 053601

\bibitem{PRAChang}
Qiao C, Tan C, Siegl J, Hu F, Niu Z, Jiang Y, Weidem\"uller M and Zhu B 2021
  {\em Phys. Rev. A\/} {\bf 103}(6) 063313

\bibitem{continuumstate}
Blow K~J, Loudon R, Phoenix S~J~D and Shepherd T~J 1990 {\em Phys. Rev. A\/}
  {\bf 42}(7) 4102--4114

\bibitem{PropagatingQED1}
Wang Y, Min\'a\ifmmode~\check{r}\else \v{r}\fi{} J~c~v, Sheridan L and Scarani
  V 2011 {\em Phys. Rev. A\/} {\bf 83}(6) 063842

\bibitem{Mollow}
Mollow B~R 1969 {\em Phys. Rev.\/} {\bf 188}(5) 1969--1975

\bibitem{collectiveDstate3}
Miroshnychenko Y, Poulsen U~V and M\o{}lmer K 2013 {\em Phys. Rev. A\/} {\bf
  87}(2) 023821

\bibitem{collectiveDstate2}
Porras D and Cirac J~I 2008 {\em Phys. Rev. A\/} {\bf 78}(5) 053816

\bibitem{collectiveDstate1}
Lehmberg R~H 1970 {\em Phys. Rev. A\/} {\bf 2}(3) 883--888

\bibitem{SuperatomDarkstate1}
Ara\'ujo M~O, Kre\ifmmode \check{s}\else \v{s}\fi{}i\ifmmode~\acute{c}\else
  \'{c}\fi{} I, Kaiser R and Guerin W 2016 {\em Phys. Rev. Lett.\/} {\bf
  117}(7) 073002

\bibitem{SuperatomDarkstate2}
Roof S~J, Kemp K~J, Havey M~D and Sokolov I~M 2016 {\em Phys. Rev. Lett.\/}
  {\bf 117}(7) 073003

\bibitem{Rydberglifetime}
Kunze S, Hohmann R, Kluge H~J, Lantzsch J, Monz L, Stenner J, Stratmann K,
  Wendt K and Zimmer K 1993 {\em Zeitschrift f{\"u}r Physik D Atoms, Molecules
  and Clusters\/} {\bf 27} 111--114

\bibitem{EITsuperatom}
Petrosyan D, Otterbach J and Fleischhauer M 2011 {\em Phys. Rev. Lett.\/} {\bf
  107}(21) 213601

\bibitem{RydbergDressingJohnson}
Johnson J~E and Rolston S~L 2010 {\em Phys. Rev. A\/} {\bf 82}(3) 033412

\end{thebibliography}

\end{document}